\documentclass[conference]{IEEEtran}

\usepackage{amsmath,amssymb,amsfonts}
\usepackage{algorithmic}
\usepackage{graphicx}
\usepackage{blindtext}
\usepackage{cite}

\usepackage[nolist]{acronym} 
\def\BibTeX{{\rm B\kern-.05em{\sc i\kern-.025em b}\kern-.08em
    T\kern-.1667em\lower.7ex\hbox{E}\kern-.125emX}}
    
\usepackage{booktabs}
\usepackage{float}

\usepackage{siunitx}
\DeclareSIUnit{\belmilliwatt}{Bm}
\DeclareSIUnit{\dBm}{\deci\belmilliwatt}
\DeclareSIUnit{\dBi}{\deci\bel i}

\usepackage{pifont}
\newcommand{\cm}{\ding{51}}%

\newcommand{\squeezeup}{\vspace{-2.5mm}}
\acused{IoT}
\acused{LoRaWAN}
\acused{GPS}

\usepackage{xcolor,colortbl}

\definecolor{DDGray}{gray}{0.8}
\definecolor{DGray}{gray}{0.82}
\definecolor{Gray}{gray}{0.85}
\definecolor{LGray}{gray}{0.9}
\definecolor{LLGray}{gray}{0.95}

\newcolumntype{e}{>{\columncolor{DDGray}}c}
\newcolumntype{f}{>{\columncolor{DGray}}c}
\newcolumntype{g}{>{\columncolor{Gray}}c}
\newcolumntype{h}{>{\columncolor{LGray}}c}
\newcolumntype{i}{>{\columncolor{LLGray}}c}

\usepackage[inline,shortlabels]{enumitem}

\begin{document}

\makeatletter
\def\ps@IEEEtitlepagestyle{%
\def\@oddfoot{\parbox{\textwidth}{\footnotesize
Author's version of a paper accepted for publication in Proceedings of the 2021 IEEE 94th Vehicular Technology Conference (VTC2021-Fall). 
\\
\textcopyright{} 2021 IEEE. 
Personal use of this material is permitted.  
Permission from IEEE must be obtained for all other uses, in any current or future media, including reprinting/republishing this material for advertising or promotional purposes, creating new collective works, for resale or redistribution to servers or lists, or reuse of any copyrighted component of this work in other works.\vspace{1.2em}}
}%
}
\makeatother

\bstctlcite{IEEEexample:BSTcontrol}
\title{Path Loss in Urban LoRa Networks:\\ A Large-Scale Measurement Study}

\author{
    \IEEEauthorblockN{Michael Rademacher\IEEEauthorrefmark{1}, Hendrik Linka\IEEEauthorrefmark{3}, Thorsten Horstmann\IEEEauthorrefmark{3}, Martin Henze\IEEEauthorrefmark{1}}  
    \IEEEauthorblockA{\IEEEauthorrefmark{1}Fraunhofer FKIE, Bonn, Germany, \{firstname.lastname\}@fkie.fraunhofer.de}  
    \IEEEauthorblockA{\IEEEauthorrefmark{3}Hochschule Bonn Rhein-Sieg, Sankt Augustin, Germany, \{firstname.lastname\}@inf.h-brs.de}
}

\maketitle

\begin{abstract}
Urban LoRa networks promise to provide a cost-efficient and scalable communication backbone for smart cities.
One core challenge in rolling out and operating these networks is radio network planning, i.e., precise predictions about possible new locations and their impact on network coverage.
Path loss models aid in this task, but evaluating and comparing different models requires a sufficiently large set of high-quality received packet power samples.
In this paper, we report on a corresponding large-scale measurement study covering an urban area of \SI{200}{\kilo\meter\squared} over a period of 230 days using sensors deployed on garbage trucks, resulting in more than 112 thousand high-quality samples for received packet power.
Using this data, we compare eleven previously proposed path loss models and additionally provide new coefficients for the \acl{Ld} model.
Our results reveal that the \acl{Ld} model and other well-known empirical models such as Okumura or Winner+ provide reasonable estimations in an urban environment, and terrain based models such as ITM or ITWOM have no advantages.
In addition, we derive estimations for the needed sample size in similar measurement campaigns.
To stimulate further research in this direction, we make all our data publicly available.
\end{abstract}

\maketitle

\section{Introduction}\label{sec:introduction}\label{sec:intro}

With smart cities moving towards reality, the possibility to connect and gather data from various assets enables municipalities and their associated utility companies to handle more complex tasks efficiently and effectively~\cite{zhang2017security}. 
For example, a timely and fine-grained collection of electricity, water, or gas consumption is imperative to further improve and control supply.
Likewise, a well planed smart city provides real benefits for its residents, ranging from detecting nearly full trashcans to identifying public green areas which need irrigation.

The most fundamental requirement for a smart city is a communication network connecting assets and things.
Wireless communication technologies are widely preferred due to (i) the lack of wired communication paths (i.e., copper or fiber) and (ii) their cost-effectiveness and better scalability.
\ac{LoRaWAN} is becoming a popular choice for smart city networks due to its long-lasting battery power devices, large cell sizes, usage of a license-exempt band and therefore independence from life-cycles of \acp{MNO}, and possible reduction of costs~\cite{trackchairpaper, yu2017study}.
However, for municipalities, setting up and operating such networks is challenging due to a lack of expertise and experience.
In particular, we identify the following issues:
\begin{enumerate*}[1)]
\item insufficient protection of devices and transmissions against tampering and misuse, 
\item difficulties of transitioning from well-functioning individual projects to professionalized mass roll-out with automated commissioning and management, and
\item non-optimal radio network planning.
\end{enumerate*}

In this work, we specifically address the issue of non-optimal radio network planning for LoRa networks. 
Specifically, our goal is to aid precise prediction about possible new locations and their impact on network coverage.
While \acp{MNO} have developed a broad knowledge of \ac{PL} models and can therefore predict coverage of their cellular technologies, this knowledge is often not public domain and in particular not available to municipalities.
To address this issue, we cooperated with the administration of a large city and associated utility companies to conduct a large-scale city-wide measurement study and evaluated several \ac{PL} models for their suitability for LoRa networks.
Our contributions are:
\begin{enumerate}
\item We survey related work on measuring LoRa networks and \ac{PL} models to derive the foundation and motivation for our measurement campaign (Section~\ref{sec:relatedWork}). 
\item We design and perform a large-scale city-wide LoRa measurement study in the city of Bonn, Germany. We gathered more than 112.000 samples over a \SI{200}{\kilo\meter\squared} urban environment using 9 different LoRa \acp{GW} and our own sensors deployed on 4 garbage trucks during a period of 230 days (Section~\ref{sec:setup}), exceeding the scale of previous work in various dimensions. 
\item We show that the \acl{Ld} model and other well-known empirical models such as Okumura or Winner+ provide reasonable estimations in an urban environment, while terrain-based models such as ITM or ITWOM have no advantage even with LiDAR terrain data (Section~\ref{sec:results}).
\item We evaluate the progression of the \acl{Ld} model coefficients and show a noticeable influence of including long-distance links, while sample sizes of above 20 thousand are needed for reliable conclusions (Section~\ref{sec:results}).
\end{enumerate}

To enable reproducibility of our results and stimulate further research, we have made all developments (i.e., schematics and software of the sensor, implementation of \ac{PL} models) and data (i.e., gathered samples) publicly available~\cite{repo-dump}. %
\section{Campaigns and Path loss Models}\label{sec:relatedWork}\label{sec:soa}

\begin{table*}[ht]
\centering
\caption{Overview of related work: Previous measurement campaigns and the evaluated \ac{PL} models}
\label{tab:overview-previous-work}
\setlength{\tabcolsep}{1.5pt}

\setlength{\aboverulesep}{0pt}
\setlength{\belowrulesep}{0pt}

\begin{tabular}{cccccccfggghhhhhhhhhii}
\multicolumn{7}{c}{\textbf{Campaign}} &  \multicolumn{1}{e}{} & \multicolumn{14}{e}{\textbf{Path loss models}} \\ \toprule
\multicolumn{7}{c}{\textbf{}} & \multicolumn{1}{f}{\textbf{}} & \multicolumn{3}{g}{\textbf{\acl{Ld}}} & \multicolumn{9}{h}{\textbf{Empirical models}} & \multicolumn{2}{i}{\textbf{Terrain}} \\ \midrule
\textbf{Paper} & \textbf{Year} & \textbf{Urban} & \textbf{Sub-Urban} & \textbf{Rural} & \textbf{Samples} & & \textbf{FSPL} & \textbf{\cite{petajajarvi2015}} & \textbf{\cite{Joerke2017}}  & \textbf{\cite{callebaut2020}} & \textbf{\cite{chall2019}} & \textbf{ITU-A} & \textbf{ITU-R} & \textbf{Winner+} & \textbf{3GPP} & \textbf{Okumura} & \textbf{COST} & \textbf{Egli} & \textbf{EC33} & \textbf{ITM} & \textbf{ITWOM}\\ \toprule
\cite{petajajarvi2015} & 2015 &\cm& & & 7012 & &\cm&\cm& & & & & & & & & & & & & \\ \midrule
\cite{Joerke2017} & 2017 &\cm& & & 546 & &\cm&\cm&\cm& & &\cm& &\cm&\cm&\cm& & & & & \\ \midrule
\cite{sancheziborra2018} & 2018 &\cm&\cm&\cm& N/A & & & & & & & & & & &\cm& & & & & \\ \midrule
\cite{linka2018} & 2018 & &\cm&\cm& 7564 & &\cm&\cm&\cm& & & & & & & & & & &\cm& \\ \midrule
\cite{delCampo2018} & 2018 &\cm&\cm& & \textless 600 & &\cm& & & & & & & & & & & & & & \\ \midrule
\cite{chall2019} & 2019 &\cm&\cm&\cm& 7000 & &\cm& & & & \cm & & & &\cm&\cm&\cm& & & & \\ \midrule
\cite{bezerra2019} & 2019 &\cm& & & N/A & & & & & & & & & & &\cm& & & &\cm&\cm\\ \midrule
\cite{callebaut2020} & 2020 &\cm&\cm&\cm& 38,146 & & &\cm& & \cm& & & & & &\cm&\cm& & & & \\ \midrule
\cite{ingabire2020} & 2020 &\cm& & & N/A & & & & & & & &\cm& & &\cm&\cm& & & & \\ \midrule
This & 2021 &\cm& & & 112,404 & &\cm&\cm&\cm&\cm&\cm& & & \cm& &\cm&\cm&\cm&\cm&\cm&\cm\\ \bottomrule 
\end{tabular}
\renewcommand{\arraystretch}{1.0}
\begin{tabular}{cccccccccccccccccccccc}
\multicolumn{22}{l}{
FSPL=Free Space Path Loss;
\cite{petajajarvi2015, Joerke2017,callebaut2020}=\acl{Ld} model\cite{molisch2012wireless};
ITU-A=ITU Advanced (UMa NLOS)\cite{itu-advanced,molisch2012wireless};
ITU-R=ITU-R M.1225\cite{itu-1225,molisch2012wireless}; 
} \\
\multicolumn{22}{l}{
Winner+=Winner+(UMa NLOS)\cite{winnerplus}; 
3GPP=3GPP Spatial (Urban Macro)\cite{3GPPSpatial, molisch2012wireless};
Okumura=Okumura Hata\cite{molisch2012wireless}; 
COST=COST231/Cost Hata\cite{molisch2012wireless};
} \\
\multicolumn{21}{l}{
Egli=Egli Model\cite{egli};
ECC33=ECC33 (ITU-R P.529)\cite{itu-529};
ITM=Longley Rice Model\cite{itm};
ITWOM=Longley-Rice and ITU-P.1546 Combined\cite{itwom}
} 
\end{tabular}
\vspace{-1.5em}
\end{table*}
Path loss (\ac{PL}) models are an important asset for radio network planning, e.g., when making prediction about possible new locations and their impact on network coverage.
To identify open issues and derive important design decisions for our own campaign, we briefly highlight the most important aspects of previous work. 
We apply a taxonomy to previously evaluated \ac{PL} models and show that individual campaigns only addressed a subset of these possible models. 
In addition, we discuss certain aspects of previous campaigns such as sample size, post-processing of samples, and lack of verification of \ac{GPS} and \ac{RPP} values. 

We survey previous work on real-world measurement campaigns and an associated evaluation of different \ac{PL} models for LoRa networks operating at $\approx 868~\si{\mega\hertz}$. 
Our comparison in Table~\ref{tab:overview-previous-work} shows numerous models (14) which have been evaluated in different measurement campaigns and environments (urban, suburban, or rural).
In this work, we apply a taxonomy of four different groups to categorize \ac{PL} models.
A complete description of all different \ac{PL} model is out of scope, however, below Table~\ref{tab:overview-previous-work}, we provide references to all models.

The first group consists of the well-known \emph{\ac{FSPL}} model~\cite{molisch2012wireless}, which is often used as a baseline~\cite{petajajarvi2015,Joerke2017,chall2019,callebaut2020, sancheziborra2018,linka2018,delCampo2018, bezerra2019,ingabire2020}. 
The second group consist of the \emph{\ac{LDPL}} model~\cite{molisch2012wireless} (one-slop model), defined as 
\begin{equation} \label{eq:ldpl}
PL(d)[dB] = 10\cdot n \cdot log_{10}(\frac{d}{d_0}) + PL_{d0} + \chi_{\sigma}
\end{equation}
In this simple model, the parameters are derived from measured samples using a curve fitting method~\cite{molisch2012wireless}. The parameter $n$ accounts for the slope, $PL_{d0}$ is the intercept and $\chi_{\sigma}$ is a zero-mean Gaussian random variable to account for the shadow fading. Based on their individual measurements campaigns, the authors in~\cite{petajajarvi2015, Joerke2017,chall2019,callebaut2020} derived different values for these parameters. %
For example, for the city of Oulu (Finland), the authors determined $n=2.65$ and $PL_{d0} = 132.25$~\cite{petajajarvi2015}. 
In~\cite{chall2019}, different antenna heights of the sensor are considered as an additional parameter to the \ac{LDPL} model and~\cite{callebaut2020} extends this model to a dual slope model.

The third group in our taxonomy are \emph{common empirical} \ac{PL} models, known from other wireless systems operating at similar frequency bands and environments. 
While these models are also based on extrapolated measurement data, they typically include additional factors such as the height of the \ac{GW} and sensor or corrections factors for different environments.
Prominent choices are the Okumura Hata~\cite{molisch2012wireless,Joerke2017, chall2019, bezerra2019, callebaut2020, ingabire2020} and the Cost Hata~\cite{molisch2012wireless,ingabire2020, callebaut2020, bezerra2019} model. 

The last group consists of models which need \emph{terrain} profiles, i.e., the ITM (Longley Rice Model~\cite{itm}) model and its successor the ITWOM (Longley-Rice and ITU-P.1546 combined~\cite{itwom}) model. 
Terrain profiles provide the possibility to include basic estimations for diffraction on obstacles along the communication path. 
Only a small subset of previous work includes these groups of models, likely, due to the increased complexity resulting from topographic data.

When comparing previous measurements campaigns, we find that these have been conducted in different \emph{environments}, with a strong focus to urban environments due to potential applications for smart cities and a general higher asset density. 
Furthermore, one important characteristic when comparing different measurement studies is the \emph{sample size} (i.e., the quantity of unique \ac{RPP} measurements). 
Sample sizes in previous work are rather small~\cite{petajajarvi2015}, ranging from 546~\cite{Joerke2017} to 38,146~\cite{callebaut2020}, though some campaigns do not explicit report sample sizes~\cite{sancheziborra2018, bezerra2019, ingabire2020}. 
Obtaining samples is time-consuming due to the relative low throughput (kb/s) and regulatory restrictions: %
To measure samples for the maximum possible range, a \ac{SF} (the modulation of LoRa) of 12 is needed. 
However, even for very small payloads (e.g., \ac{GPS} coordinates) only a few transmissions per hour (below 10) are allowed~\cite{loraScale}. 

One important aspect of measurement campaigns is the post-processing of measurement samples~\cite{callebaut2020}.
Previous work argues, that a weighted fitting of samples is needed ``to avoid that distances with a high number of samples have an excessive impact on the fitting''~\cite{callebaut2020}.
However, to the best of knowledge, previous work lacks further important discussions regarding post-processing:
First, the accuracy of \ac{GPS} coordinates is important especially for short distances below \SI{1}{\kilo\meter}. 
Second, when using \ac{COTS} LoRa modems, the correctness of the reported \ac{RPP} values should be validated as it directly influences results.

\noindent\textbf{Motivation for this work.}
Based on our discussion of related work, we identify the following shortcomings that motivate our work:
\begin{enumerate*}[1)]
\item Previous work relied on a sample size in the order of tens of thousands.
However, it is unclear if this is sufficient for statistically meaningful results~\cite{petajajarvi2015}.
\item Neither data (i.e., raw measurement samples) nor implementations of \ac{PL} models have been made available alongside previous publications.
It is thus impossible to verify results or evaluate additional \ac{PL} models.
\item The accuracy and correctness of \ac{GPS}/\ac{RPP} have not been discussed. It remains unclear if possible inaccuracies influenced results.
\item Previous publications each evaluated only a small, individual subset of \ac{PL} models with divergent results.
Therefore, we are unable to draw conclusions about the suitability of individual models for smart cities.
\end{enumerate*}

With our work, we set out to address these shortcomings and lay a foundation for further research on \ac{PL} models for LoRa networks.
To this end, we compare various \ac{PL} models in a carefully designed large-scale measurement study. %
\section{Design of the Experiment}\label{sec:setup}
In this section we describe the design and execution of our measurement campaign. Additional materials such as a documentation about the sensor, further validations and all measurement samples are available in public repositories~\cite{repo-dump}. 

\begin{figure}[t]
\centering
\includegraphics[width=\columnwidth]{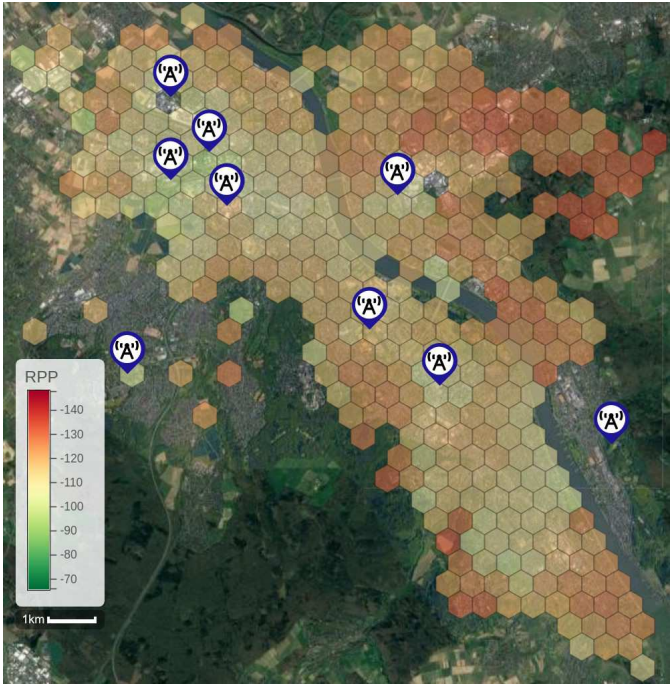}
\caption{Location of the \acp{GW} in the city of Bonn.}
\label{fig:map}
\end{figure}

\noindent\textbf{Environment.} The measurement campaign was conducted in the federal city of Bonn, Germany, a typical urban environment with tall- and medium-sized buildings with 330,000 inhabitants.
The Rhine flows through the city and the topography is flat (cf.\ Fig.~\ref{fig:map}).
The core components of this campaign are stationary LoRa \acp{GW} and mobile \ac{GPS} sensors.
Overall, 9 different LoRa \acp{GW} were distributed in the city and 4 mobile \ac{GPS} sensors generated the desired samples from all over the city area.  
The locations of the \acp{GW} were chosen based on the availability of poles and roofs.

\noindent\textbf{Hardware.}
Different \ac{GW} models from the company Kerlink were used, each equipped with a Kerlink \SI{3}{\dBi} omnidirectional antenna. 
We measured the uplink \ac{RPP} from the sensors to the \acp{GW} instead of evaluating the downlink since \begin{enumerate*}[(i)]
\item the network was already operational with other sensors from the municipality,
\item we wanted to avoid duty cycle limitations at the \acp{GW}, and
\item the sensors should be as simple as possible without the need to store the gathered data.  
\end{enumerate*}

\begin{figure}[t]
\centering
\includegraphics[width=\columnwidth]{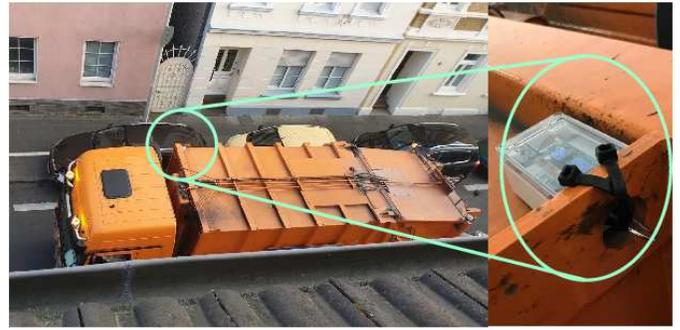}
\caption{Sensor mounted on garbage truck.}
\squeezeup
\label{fig:sensor}
\end{figure}

Our most important goal was to get a significant amount of measurement samples from all parts of the city.
Therefore, we needed vehicles that travel in the city over a long period of time.
We contacted the city's garbage collection service, which supported us with four of their trucks.
As shown in Fig.~\ref{fig:sensor}, our sensors were placed on the roof of these trucks and were thus part of their daily routes.
After initial testing, we found that the roof of a truck is a harsh environment.
Consequently, critical factors such as vibrations, water, and humidity had to be considered in the design of our sensor.

We developed our own sensor since we were unable to find a suitable one on the market. 
In addition, having control over hardware and software provides us with the possibility to evaluate more parameters of a sensor.
Our sensor is made of a processing unit, a LoRa modem, an accelerometer, a high-quality \ac{GPS} chipset, and a battery.
The main processing unit of our sensor is an ESP32 Wrover B due its low power consumption. 
The LoRa modem is a RFM95W with \SI{14}{\dBm} output power, verified using a spectrum analyzer in our laboratory. 
The antenna is placed in the cover of the enclosure to minimize the influence of the surrounding metal. 
We used a Quectel L-80R as a GPS receiver module.
Two replaceable EVE ER34615 (Lithium-thionyl Chloride) were used as power supply, in contrast to lithium polymer batteries also suitable for high temperature changes.
A complete description of the sensor, including the schematics, is provided in~\cite{repo-dump}.  

\noindent\textbf{Verification of the reported \ac{RPP}.} The \ac{RPP} measured at the various \acp{GW} is the crucial parameter for our comparison of \ac{PL} models.
Therefore, to determine the correctness of the reported \ac{RPP} values, we conducted a laboratory experiment with one of the \acp{GW} and one of our sensors before we started our measurement campaign.
We connected both antenna ports using high-frequency-cables, added additional shielding, four 80~dB static attenuators, and a dynamic attenuator.
Using the dynamic attenuator, we can adapt the attenuation without changing the overall experiment.
We transmitted 100 packets for five different attenuation values of the dynamic attenuator and obtained the \ac{RPP} at the \ac{GW}.

\begin{figure}[!t]
\centering
\includegraphics[width=\columnwidth]{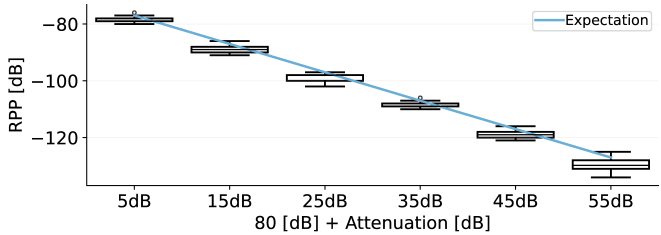}%
\caption{Linear \ac{RPP} decrease with increasing attenuation. Measured \ac{RPP} values correspond well to a simple link budget calculation.}%
\label{fig:gwrssitest}
\end{figure}

The results shown in Fig.~\ref{fig:gwrssitest} reveal that the reported \ac{RPP} decreases accordingly with an increasing attenuation, as expected. 
In addition, even the absolute values correspond to a simple link-budget estimation\footnote{Output power - static cable/connector loss - static attenuation - dynamic attenuation = \ac{RPP} $\equiv$ \SI{14}{\dBm} - \SI{6}{\deci\bel} - \SI{80}{\deci\bel} - $x$~\si{\deci\bel} = \ac{RPP}.}.
A complete documentation of our verification is provided in~\cite{repo-dump}.

\noindent\textbf{Measurement campaign.} Finally, we mounted one sensor on each of the four garbage trucks and started the measurement campaign over a period of 230 days.
If a truck (detected by the accelerometer of our sensor) moves, the \ac{GPS} is switched on.
As soon as the location has been determined, a LoRa packet with the determined \ac{GPS} coordinates is sent to all surrounding \acp{GW} with a fixed \ac{SF} of 12 (disabled \ac{ADR} function) in \ac{ABP} mode.
The used bandwidth was fixed to \SI{125}{\kilo\hertz} at a coding rate of 4/5.
Each \ac{GW} which receives a packet first determines the \ac{RPP} and afterwards forwards the packet (including the \ac{RPP}) to our logging database.
Due to our measurement setup, lost packets (i.e.\ due to collisions or insufficient signal strength) are not reported and can only be determined if at least one of the other \acp{GW} received the packet.
Overall, our measurement campaign covered more than \SI{200}{\km^{2}} urban environment and we obtained $175,492$ individual measurement samples.

\noindent\textbf{Post-processing of measurement samples.}
By investigating the raw measurements samples, we found that the accuracy of the \ac{GPS} needs to be addressed before the evaluation of different \ac{PL} models.
Due to the resulting high uncertainty of the location, we removed measurements with less than 5 locked GPS satellites ($\approx 8\%$). 
In a second step, the remaining inaccuracy of GPS measurements was compensated by mapping locations to the nearest street using the ``Open Source Routing Machine'' (OSRM) service~\cite{osrm}. 
Measurements where the offset between measured and expected street was more than 20 meters apart were also filtered out ($\approx 22\%$). 
Furthermore, measurements with an altitude higher than the highest point of the city were sorted out ($\approx 1\%$).
An indication for the resulting high accuracy of the dataset is the final filtering of all measurements with a path loss lower than the \ac{FSPL}\footnote{Physically impossible and therefore directly related to erroneous locations.}, which applied only to below 0.5\% of all data.
After post-processing, a total number of 112,372 samples remain. We did not yet evaluate the influence of a weighted fitting of samples as proposed in~\cite{callebaut2020}.
A transmitted packet of a sensor was received on average by 2.3 \acp{GW}. The maximum distance between sensor and \ac{GW} was almost \SI{13}{\km}, on average it was about \SI{3}{\km}.

\section{Results and Analysis}\label{sec:results}

Using our gathered data, we performed various analyses, including important insights for the network operator, a curve-fitting of the \acf{LDPL} model and the progression of the coefficients, the comparison of various previously proposed models (cf.\ Table~\ref{tab:overview-previous-work}), and an estimation for the needed sample population in such measurement campaigns.

\begin{figure}[t]
	\centering
	\includegraphics[width=0.95\columnwidth]{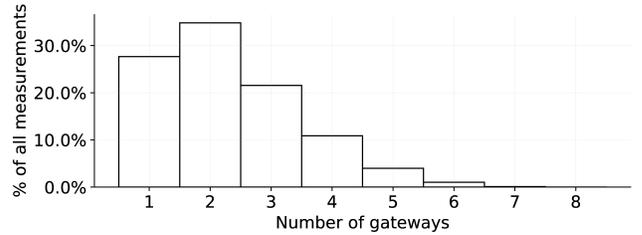}%
	\caption{Our analysis of the number of gateways receiving a certain packet shows that more than 70\% of packets are received by more than one gateway.}%
	\label{fig:gw_count}
\end{figure}

\noindent\textbf{General insights.} 
Important for the municipality is a general assessment of the network quality. As visualized in Fig.~\ref{fig:gw_count}, around 72\% of all packets were received by at least two \acp{GW}, showing good coverage of the city and redundancy of infrastructure. 

Considering the strongest received \ac{RPP} among multiple \acp{GW} from a single packet, 80\% of these packets could have been transmitted with the best modulation (SF 7). 
This would drastically improve scalability when considering techniques such as \ac{ADR} \cite{loraScale}. 
However, this would also decrease the amount of \acp{GW} which can decode a packet and therefore reduce the redundancy observed in Fig.~\ref{fig:gw_count}. This is a trade-off which needs further investigation. 

\begin{figure}[b]
	\centering
	\includegraphics[width=\columnwidth]{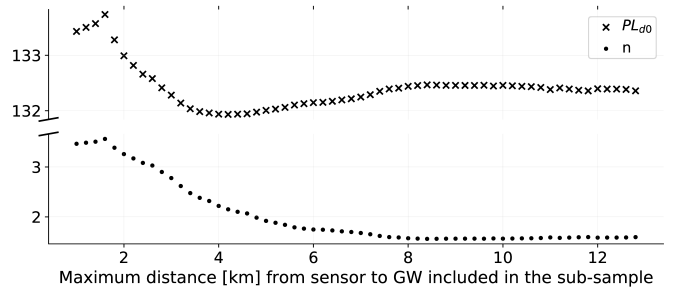}%
	\caption{Progression of the \acl{LDPL} model parameters with different maximum \ac{GW} distances in the sub-sample. Including long-distance measurements noticeably influences the exponent.}%
	\label{fig:curve_fit_distance}
\end{figure}

\noindent\textbf{\acl{LDPL} model in the city of Bonn.} Similar to previous work~\cite{petajajarvi2015,Joerke2017,callebaut2020}, we use the \ac{LDPL} model (described in Equation~\ref{eq:ldpl}) to provide an empirical representation of our measurement campaign by fitting our samples to this model.
Before the fitting, all samples were grouped by distance with an accuracy of \SI{10}{\meter} and the mean was taken for each group. 
The fitting yields an exponent of $n=1.58$ and a reference loss of $PL_{1km}=132.41$. The low value for the exponent (compared to \cite{petajajarvi2015, Joerke2017, callebaut2020}) reflects the varying topography of Bonn and in particular the long-distance measurements over several kilometers.
The influence of these long-distance measurements is additionally investigated in Fig.~\ref{fig:curve_fit_distance}. This Figure visualizes the parameters of the \ac{LDPL} model for different sub-samples of our campaign. 
When only short distances are included in a sub-sample, shadowing of buildings has a noticeable impact (high $n$).
If additionally long-distance measurements are considered, the exponent decreases and finally settles at $\approx 1.6$. 
This reveals that the \ac{LDPL} model highly depends on the maximum distance of measurements samples used for the fitting process. Therefore, researchers should always check the underlying link-distances used to circumvent using a \ac{LDPL} model out-of bounds.   

\begin{figure}[t]
	\centering
	\includegraphics[width=\columnwidth]{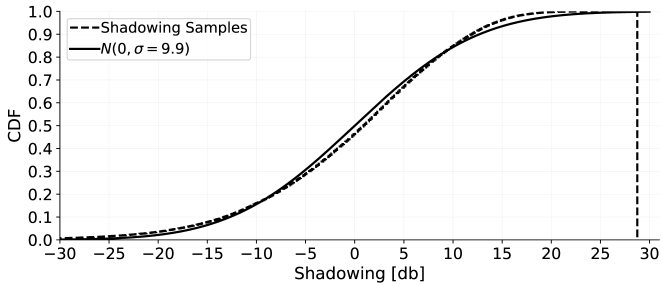}%
	\caption{\ac{ECDF} due to shadowing. $\chi_{\sigma}$ in Eq.~\ref{eq:ldpl}. The distribution follows a normal distribution with zero mean and a standard deviation of $\sigma=9.9$.}%
	\squeezeup
	\label{fig:curve_fit_cdf}
\end{figure}

The shadowing distribution is important to assess the variance at uniform distances. Fig.~\ref{fig:curve_fit_cdf} plots the \ac{ECDF} of the shadow fading samples and shows a well-fitting normal distribution with zero mean and a standard deviation of $\sigma=9.9$. 
This reflects the heterogeneity between obstructions and reflections due to urban development and sub-urban topology.

\noindent\textbf{Comparing models and real-world measurements.} 
Using our large-scale measurements, we analyze the applicability of different \ac{PL} models (cf.\ Section~\ref{sec:relatedWork}) in our urban scenario. 
To this end, each model was parameterized with all necessary information, such as antenna heights, climate factors, or building density.
In addition, for the terrain models (ITM and ITWOM), a digital elevation model (DEM) was build based on public available precise (accuracy $<$ \SI{1}{\meter}) LiDAR data~\cite{lidar}. 
To the best of our knowledge, this is the first time such precise topography data has used been to evaluate urban LoRa networks.
For most calculations, we relied on \texttt{Signal Server}~\cite{SignalServer} and implemented the remaining ones by our own. 
Our parametrization of \texttt{Signal Server} and our implementations are available at \cite{repo-dump}. 
All models produce an estimate of the \ac{PL} from the different \acp{GW} to the locations of our mobile sensors. 
The estimated \ac{PL} is afterwards included in a link budget to compare these estimates to the measured samples. 
The prediction error by a model $m$ for a sample $s$, is given by the difference between the measured and predicted \ac{RPP}: $\epsilon_{m,s} = RPP_{measured} - RPP_{predicted}$. 
To quantify the performance of the models, we used the Root Mean Square Error (RMSE).

\begin{table}[t]
\caption{RMSE for different path loss models vs. our data-set.}
\centering
\setlength{\tabcolsep}{0.7pt}
\setlength{\aboverulesep}{0.5pt}
\setlength{\belowrulesep}{0.5pt}
\begin{tabular}{ccccccc}
\rowcolor{Gray} 
& \multicolumn{6}{c}{\textbf{Baseline and \acl{Ld} models}} \\ \toprule
& \textbf{FSPL} & \textbf{Oulu\cite{petajajarvi2015}} & \textbf{Dortmund\cite{Joerke2017}} & \textbf{Ghent\cite{callebaut2020}} & \textbf{This work} & \\
\textbf{RMSE [\si{\dBm}]} & 40.85 & 10.17 & 11.07 & 31.05 & 9.9 & \\
\rowcolor{LGray} 
& \multicolumn{6}{c}{\textbf{Empirical models}} \\ \toprule
& \textbf{Beirut~\cite{chall2019}} & \textbf{Winner+} & \textbf{Okumura} & \textbf{COST} & \textbf{Egli} & \textbf{EC33} \\
\textbf{RMSE [\si{\dBm}]} & 16.38 & 12.11 & 10.92 & 12.51 & 23.25 & 25.24 \\
\rowcolor{LLGray} 
& \multicolumn{6}{c}{\textbf{Terrain models}} \\ \toprule
& \textbf{ITM} & \textbf{ITWOM} & & & & \\
\textbf{RMSE [\si{\dBm}]} & 23.25 & 21.81 &  &  &  & \\
\end{tabular}
\label{tab:res-rmse}
\squeezeup
\end{table}

Table~\ref{tab:res-rmse} shows the RMSE when applying previous proposed \ac{PL} models to our measurements. 
The best performing models achieve an RMSE of \SI{10}{\dBm}, making them suitable for real-world network planing purposes.
Especially the \ac{LDPL} models provide good predictions. 
An exception is the \ac{LDPL} model from Ghent~\cite{callebaut2020} which provides the worst prediction of all models for our dataset. 
For the group of empirical models, Winner+ and Cost-Hata have low error rates as well. 
This is expected, as they are all based on real-world measurements in urban environments. 
Interestingly, the terrain-based models ITM and ITWOM performed significantly worse. 
Further research is needed to determine the precise reasons, however, diffractions are probably overestimated and the more dominant reflections on buildings are not taken into account~\cite{itm, itwom}. 

\begin{figure}[b]
	\centering
	\includegraphics[width=\columnwidth]{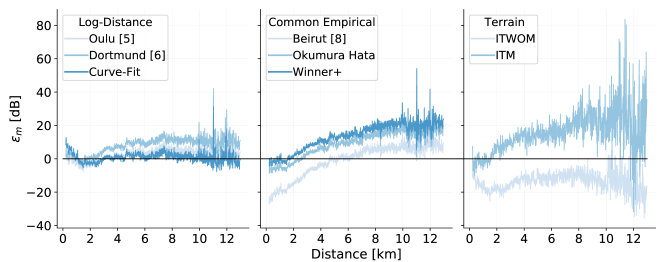}%
	\caption{Difference between measured and predicted \ac{RPP} for varying distance. Overestimation of \ac{RPP} below and underestimation of \ac{RPP} above the abscissa.}%
	\label{fig:skewness_distance}
\end{figure}

While the RMSE is suitable for a general quality assessment of \ac{PL} models this simple metric neglects possible systematic deviations.
Therefore, the next analysis shown in Fig.~\ref{fig:skewness_distance}, considers deviations of the models over varying distance using the difference between measured and predicted \ac{RPP} ($\epsilon$). 
Here, the empirical models (Beirut~\cite{chall2019}, Okumura Hata and Winner+) show a similar trend, as they overestimate the \ac{RPP} for short distances and underestimate it for longer links. 
The \ac{LDPL} models tend to slightly underestimate the \ac{RPP}. 
ITM and ITWOM make opposing predictions. ITM underestimate the \ac{RPP} whereas ITWOM overestimate it and both show a high variance due to the inclusion of different diffraction situations.
The practical interpretation of these results is highly application dependent, as it is difficult to know in advance whether overestimates or underestimates are more harmful.

\begin{figure}[t]
	\centering
	\includegraphics[width=\columnwidth]{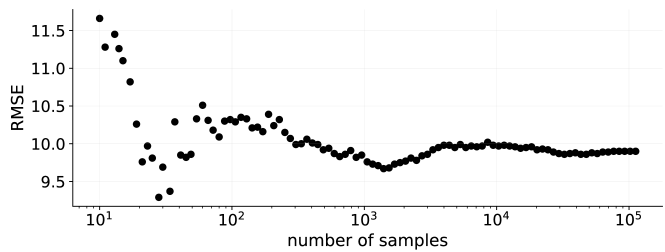}%
	\caption{Progression of the RMSE depending on the number of samples included in the fitting of our \acl{LDPL} model. A significant number of samples are needed before the RMSE is stable.}%
	\squeezeup
	\label{fig:curve_fit_rmse_progression}
\end{figure}

In our measurement, we collected a significant number of samples to compare the different \ac{PL} models and parametrize the \ac{LDPL} model with statistically meaningful data. 
To this end, Fig.~\ref{fig:curve_fit_rmse_progression} investigates the number of samples required for the curve-fitting procedure to converge, i.e., how many measurement samples are needed to derive meaningful results.
Here, the samples were randomly chosen as a subset of all measurements samples. 
Although nearly all related work uses sample size of less than 10 thousand, approximately 20-30 thousand samples are needed before the RMSE of the \ac{LDPL} model becomes stable.
Especially when sample sizes below 1000 are used for fitting and comparison (cf.\ Table~\ref{tab:overview-previous-work}), authors should discuss this aspect more critically.

\section{Conclusion and Future Work}\label{sec:conclusion}
In this work, we have presented the design and evaluation of a large-scale measurement study to quantify the path loss in urban LoRa networks.
This work exceeds previous efforts in this field in various dimensions (sample size, duration, comparison of different \acl{PL} models).
We found that simple \acl{Ld} or empirical \acl{PL} models provide good estimations in an urban environment, while more complex terrain-based models (ITM or ITWOM) do not increase prediction quality, even with precise LiDAR data.
In addition, our large sample size leads to additional insights over previous work:
Common empirical models overestimate the \ac{RPP} for short distances and underestimate it for long distances.
We show that 20-30 thousand samples are needed to conduct a reliable fitting process for the \acl{LDPL} model and that long-distance links significantly influence the results. 

\noindent\textbf{Future work.}
Our publicly available dataset~\cite{repo-dump} provides ample opportunities for further investigations:
\begin{enumerate*}[1)]
\item As an individual packet ID is included in our dataset, phenomena such as the average packet loss can be studied.
\item We focused on \ac{PL} models, which have been used in previous work. In the future, new models may emerge and an investigation of these models using our dataset should be a straight forward task.
\item By linking the \ac{RPP} to the modulation (SF), a more precise conclusion on the scalability of LoRa networks in urban environments could be drawn.
\item Using alternative models which use LiDAR topography data could lead to significantly better results, with ray-tracing being one promising approach.
\end{enumerate*}
With our work, we lay the foundation for such interesting investigations and follow-up work.
By making all our data and implementations publicly available~\cite{repo-dump}, we further stimulate research on the large-scale analysis of urban LoRa networks and large area wireless sensor networks in general.
Already today, our results reveal the necessity to consider a large sample size when studying different path loss models. %

\section*{acknowledgements}
\noindent The authors would like to thank ``bonnorange AöR'' for providing us with the opportunity to use their garbage trucks. 

\begin{acronym}[]
\acro{3GPP}{3rd Generation Partnership Project}
\acro{AC}{Access Category}
\acro{ACK}{acknowledgement}
\acro{AI}{Abstract Interface}
\acro{AIFS}{Arbitration Interframe Space}
\acro{AODV}{Ad hoc On-Demand Distance Vector Routing Protocol}
\acro{AP}{Access Point}
\acro{API}{application programming interface}
\acro{ARP}{Address Resolution Protocol}
\acro{ARQ}{Automatic Repeat reQuest}
\acro{AS}{Autonomous System}
\acro{ASCII}{American Standard Code for Information Interchange}
\acro{ATIS}{Alliance for Telecommunications Industry Solutions}
\acro{ATM}{Asynchronous Transfer Mode}
\acro{B.A.T.M.A.N.}{Better Approach to MANET}
\acro{BER}{Bit Error Rate}
\acro{BFS-CA}{Breadth First Search Channel Assignment}
\acro{BGP}{Border Gateway Protocol}
\acro{BPSK}{Binary Phase-Shift Keying}
\acro{BRA}{Bidrectional Routing Abstraction}
\acro{BS}{Base station}
\acro{BSSID}{Basic Service Set Identification}
\acro{BTS}{Base Transceiver Station}
\acro{CA}{Channel Assignment}
\acro{CAPEX}{capital expenditure}
\acro{CAPWAP}{Control And Provisioning of Wireless Access Points}
\acro{CARD}{Channel Assignment with Route Discovery}
\acro{CAS}{Channel Assignment Server}
\acro{CCA}{Clear Channel Assessment}
\acro{CDMA}{Code Division Multiple Access}
\acro{CF}{CompactFlash}
\acro{CIDR}{Classless Inter-Domain Routing}
\acro{CLICA}{Connected Low Interference Channel Assignment}
\acro{CLI}{Command Line Interface}
\acro{COTS}{Commercial Off-the-Shelf}
\acro{CPE}{Customer Premises Equipment}
\acro{CPU}{Central Processing Unit}
\acro{CRAHN}{Cognitive Radio Ad-Hoc Network}
\acro{CRCN}{Cognitive Radio Cellular Network}
\acro{CR}{Cognitive Radio}
\acro{CR-LDP}{Constraint-based Routing Label Distribution Protocol}
\acro{CRN}{Cognitive Radio Network}
\acro{CRSN}{Cognitive Radio Sensor Network}
\acro{CRVN}{Cognitive Radio Vehicular Network}
\acro{CSMA/CA}{Carrier Sense Multiple Access/Collision Avoidance}
\acro{CSMA}{Carrier Sense Multiple Access}
\acro{CSMA/CD}{Carrier Sense Multiple Access/Collision Detection}
\acro{CTA}{Centralized Tabu-based Algorithm}
\acro{CW}{Contention Window}
\acro{CWLAN}{Cognitive Wireless Local Area Network}
\acro{CWMN}{Cognitive Wireless Mesh Network}
\acro{DAD}{Duplicate Address Detection}
\acro{DCF}{Distributed Coordination Function}
\acro{DCiE}{Data Center Infrastructure Efficiency}
\acro{DDS}{Direct digital synthesizer}
\acro{DFS}{Dynamic Frequency Selection}
\acro{DGA}{Distributed Greedy Algorithm}
\acro{DHCP}{Dynamic Host Configuration Protocol}
\acro{DIFS}{Distributed Interframe Space}
\acro{DMesh}{Directional Mesh}
\acro{D-MICA}{Distributed Minimum Interference Channel Assignment}
\acro{DR}{Designated Router}
\acro{DSA}{Dynamic Spectrum Allocation}
\acro{DSLAM}{Digital Subscriber Line Access Multiplexer}
\acro{DSL}{Digital Subscriber Line}
\acro{DSR}{Dynamic Source Routing Protocol}
\acro{DSSS}{Direct-Sequence Spread Spectrum}
\acro{DTCP}{Dynamic Tunnel Configuration Protocol}
\acro{DVB}{Digital Video Broadcast}
\acro{DVB-H}{Digital Video Broadcast - Handheld}
\acro{DVB-RCS}{Digital Video Broadcast - Return Channel Satellite}
\acro{DVB-S2}{Digital Video Broadcast - Satellite - Second Generation}
\acro{DVB-S}{Digital Video Broadcast - Satellite}
\acro{DVB-SH}{Digital Video Broadcast - Satellite services to Handhelds}
\acro{DVB-T2}{Digital Video Broadcast - Second Generation Terrestrial}
\acro{DVB-T}{Digital Video Broadcast - Terrestrial}
\acro{E2CARA-TD}{Energy Efficient Channel Assignment and Routing Algorithm – Traffic Demands}
\acro{ECN}{Explicit Congestion Notification}
\acro{ECDF}{Empirical Cumulative Distribution Function}
\acro{EDCA}{Enhanced Distributed Coordination Access}
\acro{EDCF}{Enhanced Distributed Coordination Function}
\acro{EGP}{Exterior Gateway Protocol}
\acro{EICA}{External Interference-Aware Channel Assignment}
\acro{EIFS}{Extended Interframe Space}
\acro{EIGRP}{Enhanced Interior Gateway Routing Protocol}
\acro{EIRP}{Equivalent Isotropically Radiated Power}
\acro{EPI}{energy proportionality index}
\acro{ERO}{Explicit Route Object}
\acro{ETSI}{European Telecommunications Standards Institute}
\acro{ETT}{Expected Transmission Time}
\acro{ETX}{Expected Transmission Counts}
\acro{FCC}{Federal Communications Commission}
\acro{FCS}{Frame Check Sequence}
\acro{FDMA}{Frequency Division Multiple Access}
\acro{FEC}{Forward Error Correction}
\acro{FIFO}{First-In-First-Out}
\acro{FLOPS}{Floating-Point Operations Per Second}
\acro{FRR}{Fast Reroute}
\acro{FSL}{Free-Space Loss}
\acro{FSPL}{Free-Space Path Loss}
\acro{GAN}{Generic Access Network}
\acro{GDP}{Gross Domestic Product}
\acro{GEO}{Geosynchronous Earth Orbit}
\acro{GMPLS}{Generalized Multiprotocol Label Switching}
\acro{GPS}{Global Positioning System}
\acro{GRE}{Generic Routing Encapsulation}
\acro{GSE}{Generic Stream Encapsulation}
\acro{GSM}{Global System for Mobile Communications}
\acro{GW}{Gateway}
\acro{HAP}{High-Altitude Platform}
\acro{HCCA}{HCF controlled channel access}
\acro{HCF}{Hybrid Coordination Function}
\acro{HLR}{Home Location Register}
\acro{HOL}{Head-of-line}
\acro{HOLSR}{Hieracical Optimised Link State Routing}
\acro{HPC}{hardware performance counters}
\acro{HSLS}{Hazy-Sighted Link State Routing Protocol}
\acro{HWMP}{Hybrid Wireless Mesh Protocol}
\acro{IAX2}{Inter-Asterisk eXchange Version 2}
\acro{IBSS}{Independent Basic Service Set}
\acro{ICMP}{Internet Control Message Protocol}
\acro{ICT}{Information and Communication Technologie}
\acro{IEEE}{Institute of Electrical and Electronics Engineers}
\acro{IE}{Information Element}
\acro{IETF}{Internet Engineering Task Force}
\acro{IETF}{The Internet Engineering Task Force}
\acro{IFS}{Interframe Space}
\acro{ITU}{International Telecommunication Union}
\acro{IGP}{Interior Gateway Protocol}
\acro{IGRP}{Interior Gateway Routing Protocol}
\acro{ILP}{Integer Linear Programming}
\acro{ILS}{Iterated Local Search}
\acro{IPFIX}{IP Flow Information Export}
\acro{IP}{Internet Protocol}
\acro{IPv4}{Internet Protocol}
\acro{IPv6}{Internet Protocol, Version 6}
\acro{ISI}{Inter-symbol interference}
\acro{IS-IS}{Intermediate system to intermediate system}
\acro{ISM}{Industrial, Scientific and Medical}
\acro{ISP}{Internet Service Provider}
\acro{LAA}{Licensed-Assisted Access}
\acro{LDC}{Least Developed Countries}
\acro{LA-CA}{Load-Aware Channel Assignment}
\acro{LCOS}{LANCOM Operating System}
\acro{LDP}{Label Distribution Protocol}
\acro{Ld}{Log-distance}
\acro{LDPL}{Log-distance path loss}
\acro{LEO}{Low Earth Orbit}
\acro{LER}{Label Edge Router}
\acro{LGI}{Long Guard Interval}
\acro{LLTM}{Link Layer Tunneling Mechanism}
\acro{LMA}{Local Mobility Anchor}
\acro{LMP}{Link Management Protocol}
\acro{LoS}{Line of Sight}
\acro{LOS}{Line of Sight}
\acro{LQF}{Longest-Queue-First}
\acro{LS}{Link State}
\acro{LSP}{Label-Switched Path}
\acro{LSR}{Label-Switched Router}
\acro{LST}{Link-State-Table}
\acro{LTE}{Long Term Evolution}
\acro{LTE-M}{\ac{LTE} Machine Type Communication}
\acro{LWAPP}{Lightweight Access Point Protocol}
\acro{MAC}{Media Access Control}
\acro{MAG}{Mobile Access Gateway}
\acro{MANET}{Mobile Adhoc Network}
\acro{MBMS}{Multimedia Broadcast Multicast Service}
\acro{MCG}{multi-conflict graph}
\acro{MCI-CA}{Matroid Cardinality Intersection Channel Assignment}
\acro{MCS}{Modulation and Coding Scheme}
\acro{MDR}{MANET Designated Router}
\acro{MEO}{Medium Earth Orbit}
\acro{MICS}{Media Independent Command Service}
\acro{MIES}{Media Independent Event Service}
\acro{MIHF}{Media Independent Handover Function}
\acro{MIHF++}{Media Independent Handover Function++}
\acro{MIH}{Media Independent Handover}
\acro{MIIS}{Media Independent Information Service}
\acro{MILP}{Mixed Integer Linear Programming}
\acro{MIMF}{Media Independent Messaging Function}
\acro{MIMO}{Multiple Input Multiple Output}
\acro{MNO}{Mobile Network Operator}
\acro{MIMS}{Media Independent Messaging Service}
\acro{MIPS}{Million Instruction Per Second}
\acro{MMF}{Mobility Management Function}
\acro{MN}{Mesh Node}
\acro{MonF}{Monitoring Function}
\acro{MPDU}{MAC Protocol Data Unit}
\acro{MPEG}{Moving Picture Experts Group}
\acro{MPE}{Multi Protocol Encapsulation}
\acro{MPLCG}{Multi-Point Link Conflict Graph}
\acro{MPLS}{Multiprotocol Label Switching}
\acro{MPLS}{Multi Protocol Label Switching}
\acro{MPLS-TE}{Multi Protocol Label Switching - Traffic Engineering}
\acro{MP}{Merge Point}
\acro{MPR}{Multipoint Relay}
\acro{MR-MC WMN}{Multi-Radio Multi-Channel Wireless Mesh Network}
\acro{MSC}{Mobile-services Switching Centre}
\acro{MSDU}{MAC Service Data Unit}
\acro{MSTP}{Mobility Services Transport Protocol}
\acro{MT}{Mobile Terminal}
\acro{MTU}{Maximum Transmission Unit}
\acro{NAV}{Network Allocation Vector}
\acro{ns-3}{network simulator 3}
\acro{NBMA}{Non-broadcast Multiple Access}
\acro{NetEMU}{Network Emulator}
\acro{NLOS}{None Line of Sight}
\acro{NMEA}{National Marine Electronics Association}
\acro{NPC}{Normalized Power Consumption}
\acro{NP}{Nondeterministic Polynomial Time}
\acro{NSIS}{Next Steps in Signaling}
\acro{NTP}{Network Time Protocol Unit}
\acro{OFDMA}{Orthogonal Frequency Division Multiple Access}
\acro{OFDM}{Orthogonal Frequency Division Multiplex}
\acro{OLSR}{Optimized Link State Routing}
\acro{OPEX}{operational expenditure}
\acro{OSA}{Opportunistic Spectrum Access}
\acro{OSI}{Open Systems Interconnection}
\acro{OSPF}{Open Shortest Path First}
\acro{OSPF-TE}{Open Shortest Path First - Traffic Engineering}
\acro{OVS}{Open vSwitch}
\acro{P2MP}{Point To Multipoint}
\acro{P2P}{Point To Point}
\acro{PA}{Power amplifier}
\acro{PCE}{Path Computation Element}
\acro{PCEP}{Path Computation Element Protocol}
\acro{PCF}{Path Computation Function}
\acro{PCF}{Point Coordination Function}
\acro{PDR}{Packet Delivery Ratio}
\acro{PDV}{Packet Delay Variation}
\acro{PER}{Packet Error Rate}
\acro{PLCP}{Physical Layer Convergence Protocol}
\acro{PLL}{Phase-Locked Loop}
\acro{PL}{path loss}
\acro{PLR}{Point of Local Repair}
\acro{PMIP}{Proxy Mobile IP}
\acro{PoE}{Power over Ethernet}
\acro{PPDU}{Physical Protocol Data Unit}
\acro{PPP}{Precise Point Positioning}
\acro{PSTN}{Public Switched Telephone Network}
\acro{PTP}{Precision Time Protocol}
\acro{PUE}{Power Usage Effectiveness}
\acro{PU}{Primary User}
\acro{QAM}{Quadrature amplitude modulation}
\acro{QoS}{Quality of Service}
\acro{RAND}{Random Channel Assignment}
\acro{RERR}{Route Error}
\acro{RFC}{Request for Comments}
\acro{RIPng}{Routing Information Protocol next generation}
\acro{RIP}{Routing Information Protocol}
\acro{RPC}{Remote Procedure Call}
\acro{RPP}{Received Packet Power}
\acro{RREP}{Route Reply}
\acro{RREQ}{Route Request}
\acro{RSSI}{Received Signal Strength Indication}
\acro{RSS}{Received Signal Strength}
\acro{RSVP}{Resource ReSerVation Protocol}
\acro{RSVP-TE}{Resource ReSerVation Protocol - Traffic Engineering}
\acro{RTK}{Real Time Kinematic}
\acro{RTP}{Real-time Transport Protocol}
\acro{RTS}{Ready-To-Send}
\acro{RTT}{Round Trip Time}
\acro{SAA}{Stateless Address Autoconfiguration}
\acro{SAPOS}{Satellitenpositionierungsdienst der deutschen Landesvermessung}
\acro{SAP}{Service Access Point}
\acro{SBC}{Single-Board Computer}
\acro{SBM}{Subnetwork Bandwidth Manager}
\acro{SBR}{System zur Bestimmung des Richtungsfehlers}
\acro{SC-FDMA}{Single Carrier Frequency Division Multiple Access}
\acro{SDMA}{Space-division multiple access}
\acro{SDN}{Software Defined Networking}
\acro{SDR}{Software Defined Radio}
\acro{SDWN}{Software Defined Wireless Networks}
\acro{SENF}{Simple and Extensible Network Framework}
\acro{SGI}{Short Guard Intervall}
\acro{SIFS}{Short Interframe Space}
\acro{SINR}{Signal-to-Noise-plus-Interference Ratio}
\acro{SIP}{Session Initiation Protocol}
\acro{SISO}{Single Input Single Output}
\acro{SNR}{signal-to-noise ratio}
\acro{SONET}{Synchronous Optical Networking}
\acro{SPoF}{Single Point of Failure}
\acro{SSID}{Service Set Identifier}
\acro{STP}{Spanning Tree Protocol}
\acro{SU}{Secondary User}
\acro{TCP}{Transmission Control Protocol}
\acro{TPC}{Transmission Power Control}
\acro{TC}{Topology Control}
\acro{TDMA}{Time Division Multiple Access}
\acro{TEEER}{Telecommunications Equipment Energy Efficiency Rating}
\acro{TEER}{Telecommunications Energy Efficiency Ratio}
\acro{TE}{Traffic Engineering}
\acro{TIM}{Technology Independend Monitoring}
\acro{TLV}{Type-Length-Value}
\acro{TORA}{Temporally-Ordered Routing Algorithm}
\acro{ToS}{Type of Service}
\acro{TSFT}{Time Synchronization Function Timer}
\acro{TTL}{Time to live}
\acro{TVWS}{TV White Space}
\acro{TXOP}{Transmit opportunity}
\acro{UAV}{Unmanned Aerial Vehicle}
\acro{UDLR}{Unidirectional Link Routing}
\acro{UDL}{Unidirectional Link}
\acro{UDP}{User Datagram Protocol}
\acro{UDT}{Unidirectional Technology}
\acro{UE}{User Equipment}
\acro{UHF}{Ultra High Frequency}
\acro{UMA}{Unlicensed Mobile Access}
\acro{UMTS}{Universal Mobile Telecommunications System}
\acro{U-NII}{Unlicensed National Information Infrastructure}
\acro{UPS}{Uninterruptible Power Supply}
\acro{UP}{User Priorities}
\acro{USB}{Univeral Serial Bus}
\acro{USO}{Universal Service Obligation}
\acro{USRP}{Universal Software Radio Peripheral}
\acro{VCO}{Voltage-Controlled Oscillator}
\acro{VoIP}{Voice-over-IP}
\acro{VPN}{Virtual Prvate Network}
\acro{WBN}{Coordinated Wireless Backhaul Network}
\acro{WDS}{Wireless Distribution System}
\acro{WiBACK}{Wireless Back-Haul}
\acro{Wi-Fi}{Wireless Fidelity}
\acro{WiLD}{WiFi-based Long Distance}
\acro{WiMAX}{Worldwide Interoperability for Microwave Access}
\acro{WISPA}{Wireless Internet Service Provider Assocication}
\acro{WISP}{Wireless Internet Service Provider}
\acro{WLAN}{Wireless Local Area Network}
\acro{WLC}{Wireless LAN Controller}
\acro{WMN}{Wireless Mesh Network}
\acro{WMN}{Wireless Mesh Network}
\acro{wmSDN}{Wireless Mesh Software Defined Network}
\acro{WNIC}{Wireless Network Interface Controller}
\acro{WN}{WiBACK Node}
\acro{WRAN}{Wireless Regional Area Network}
\acro{WSN}{Wireless Sensor Network}
\acro{ZigBee}{ZigBee Alliance IEEE 802.15.4}
\acro{ZPR}{Zone Routing Protocol}

\acro{ABP}{Activation by Personalization}
\acro{ADR}{Adaptive Data Rate}
\acro{IoT}{Internet of Things}
\acro{LPWAN}{low-power wide-area network}
\acro{LoRaWAN}{Long Range Wide Area Network}
\acro{GPRS}{General Packet Radio Service}
\acro{CEP}{Circular Error Probable}
\acro{EIRP}{equivalent isotropically radiated power}
\acro{ITM}{Longley-Rice Irregular Terrain Model}
\acro{SF}{Spreading Factor}
\end{acronym}

\bibliographystyle{IEEEtran}
\bibliography{./bibabbrv.bib, ./lit.bib}

\end{document}